\documentclass[%
reprint,
superscriptaddress,
amsmath,amssymb,
prf,
longbibliography,
]{revtex4-1}
\usepackage[export]{adjustbox}
\usepackage{graphicx}% Include figure filessn
\usepackage{color}% Include figure files
\usepackage{xcolor}% Include figure files
\usepackage{wasysym}% Include figure files
\usepackage{dcolumn}% Align table columns on decimal point
\usepackage{bm}% bold math
\usepackage[mathscr]{euscript}
\graphicspath{{fig/}}
\definecolor{mygreen}{rgb}{0,0.5,0}
\definecolor{mybrown}{rgb}{0.65,0.16,0.16}
\newcommand{\colg}[1]{\textcolor{mygreen}{#1}}

\newcommand{\colb}[1]{\textcolor{blue}{#1}}
\newcommand{\colr}[1]{\textcolor{red}{#1}}
\newcommand{\colm}[1]{\textcolor{magenta}{#1}}

\newcommand{\colbr}[1]{\textcolor{mybrown}{#1}}

\def\druvec{{\delta} \mathbf{u}}
\def\dru{{\delta} u}

\def\yjmr{Y_{jm}(\urh)}
\def\ljm{\Lambda^{(p,L)}_{j,m}}
\def\tjm{\Lambda^{(p,T)}_{j,m}}
\def\vslp {{{{\xi}^L_j}(p)}}
\def\vstp {{{{\xi}^T_j}(p)}}
\def\vsdim {{{{\xi}_{j,dim}}(p)}}

\def\beq {\begin{equation}}
\def\eeq {\end{equation}}
\def\beqa {\begin{eqnarray}}
\def\eeqa {\end{eqnarray}}
\def \bnum {\begin{enumerate}}
\def \enum {\end{enumerate}}
\def\bi {\begin{itemize}}
\def\ei {\end{itemize}}
\def \bdes {\begin{description}}
\def \edes {\end{description}}
\def\meandiss {\epsilon}
\def\rel {R_{\lambda}}

\def\dho{\partial}

\def\normr {r/\eta}

\def\la {\langle}
\def\ra {\rangle}

\def\kres{{k_{max}\eta}}

\def\ur { \mathbf{r} }
\def\urmag {r }
\def\urh { \hat{\mathbf{r}} }

\def\ur { \mathbf{r} }

\def\uu { \mathbf{u} }
\def\ur { \mathbf{r} }

\def\kk {{\mathbf{k}}}

\def\ux { \mathbf{x} }

\def\uf { \mathbf{f} }

\def\normt{t/T_E}

\def\sot{SO(3)}
\def\bij{b_{ij}}
\def\iplus{I_{+}}
\def\iminus{I_{-}}
\def\td{\tilde}

\begin{document}
%\setlength{\abovedisplayskip}{3pt}
%\setlength{\belowdisplayskip}{3pt}
%\preprint{APS/123-QED}
\title{Multiscale anisotropic fluctuations in sheared turbulence with multiple states}
%%%%%%%%%%%%%%%%%%%%%%%%%%
\author{Kartik P. Iyer}
\email{kartik.iyer@nyu.edu}
\affiliation{
Department of Physics and INFN, 
University of Rome Tor Vergata,
Rome, $00133$, Italy
}
\affiliation{
Department of Mechanical Engineering, 
New York University, New York 11201, USA
}
\author{Fabio Bonaccorso}
\affiliation{
Department of Physics and INFN, 
University of Rome Tor Vergata,
Rome, $00133$, Italy
}
\author{Luca Biferale}
\affiliation{
Department of Physics and INFN, 
University of Rome Tor Vergata,
Rome, $00133$, Italy
}
\author{Federico Toschi}
\affiliation{
Department of Physics, Eindhoven University of Technology, $5600$ MB Eindhoven, The Netherlands
}
\affiliation{
IAC CNR, $00185$, Rome, Italy
}
%%%%%%%%%
\date{Postprint version of the manuscript published in Phys. Rev. Fluids, $2$, $052602$(R)$(2017)$}
\begin{abstract}
We use
%changed by KI due to Ref1 comments
%%%%%state-of-the-art direct numerical simulations to
high resolution direct numerical simulations to
study the anisotropic contents of a turbulent,
statistically homogeneous flow with random transitions
among multiple energy containing states.
%%%changed by KI for Ref1
%%%Using a novel algorithm we decompose the velocity
%%%correlation functions on different sectors
%%%of the three dimensional group of rotations, $\sot$. 
We decompose the velocity
correlation functions on different sectors
of the three dimensional group of rotations, $\sot$, using a 
high-precision quadrature. 
Scaling properties of anisotropic components of longitudinal and
%%%changed by KI due to Ref.1 comments
%%%transverse velocity fluctuations are measured with unprecedented accuracy,
transverse velocity fluctuations are accurately measured
at changing Reynolds numbers.
We show that independently of the anisotropic content of the 
energy containing eddies,
small-scale turbulent fluctuations recover {\it isotropy} and 
{\it universality}
faster than previously reported
in experimental and numerical studies.
The discrepancies are ascribed to the presence of highly
anisotropic contributions that have either been neglected 
or measured with less accuracy in the foregoing works.
Furthermore, the anomalous anisotropic scaling exponents are devoid of any sign
of saturation with increasing order.
Our study paves the way to systematically assess persistence of
anisotropy in high Reynolds number flows.
\end{abstract}
%%%%%%%%%%%%%%%%%
\maketitle
%%%%%%%%%%%%%%%%%
%%%%START%%%%%%%%%%%%%
\noindent The notion that all turbulent flows attain {\it universal} properties 
at small scales, regardless of the macroscopic details, 
has been an enduring and 
yet unproved concept 
in turbulence research \cite{K41a,Fri95,pope}.
The energy containing scales in many flows such as shear,
rotation, natural convection, thick layers, atmospheric boundary layer
and magneto-hydrodynamic flows,
are all strongly affected by anisotropic (and non-homogeneous)
effects of the extrinsic stirring
and boundary conditions, resulting in seemingly different
flow configurations
\cite{DTS97,GM15,xia11,SM13,GPSC13,CC13,RE15,MP15,JS16,SS16}.
As such, anisotropic fluctuations are always connected to
some degree of non-universality, i.e.~dependency on the empirical setup.
Can we disentangle anisotropic from
isotropic statistical contributions?
Are there any universal facets of turbulence? 
How does the relative importance of
anisotropic and isotropic fluctuations vary with turbulence intensity?
These are the questions we attempt to address.
%%%%%%%%%%%%%%%%%%%%%%%%%%%%%%%%%%%
 
%\noindent On one hand, all phenomenological turbulence theories 
%and almost all
%experimental and numerical
%data point toward the existence of a return-to-isotropy, at small enough scales
%\cite{K41a,Fri95,pope}. 
%%%%
\noindent On one hand, all phenomenological turbulence theories 
point toward a return-to-isotropy, at small enough scales
\cite{K41a,Fri95,pope}. 
On the other hand,
measurements of anisotropic contributions as functions of
scale separation has revealed persistent small-scale
anisotropy in hydrodynamical
turbulence \cite{TavCorrsin81,PS95,GW98,SW2000,WS02}, 
magneto-hydrodynamics \cite{muller03,watson04},
and passive scalar mixing \cite{War20,bp05}.
The persistence of anisotropy as reported
in Refs.~\cite{GW98,SW2000,War20},
was later reconciled with the postulate of
local isotropy as an effect of the  
existence of
anomalous scaling in both isotropic and anisotropic correlation
functions \cite{BV2001,bp05}.
\noindent In this letter, we investigate the
return-to-isotropy {\it{vs}} persistence-of-anisotropy, 
using direct numerical simulations (DNS)
of turbulent flows subject to large-scale
shear at high Reynolds numbers, $ Re \equiv u'r_f/\nu$,  where
$r_f$ denotes the typical forcing scale, $u'$ the root-mean-square velocity fluctuation
and $\nu$ is the viscosity.
%We  analyze the statistics using an {\it exact} decomposition of 
We  use an {\it exact} decomposition of 
multi-point turbulent
correlation functions in the eigen-basis of the $\sot$ group of rotations,
which is the only systematic method to disentangle isotropic from 
anisotropic contributions,
and to further distinguish among different anisotropic 
turbulent fluctuations.
However, the utility of the $\sot$ decomposition
has largely been impeded by practical difficulties in both experiments and simulations.
High-Reynolds number experiments
are beset with limitations on the set of directions that can be probed in
three-dimensional (3D) space
and consequently resort to ad-hoc curve fits to separate isotropic from anisotropic
scaling properties \cite{adsk98}.
Similarly, simulations have until now managed to perform the
$\sot$ decomposition at low Reynolds numbers only \cite{bp05}, due to 
computational bottlenecks
(see Supplemental Material at \cite{SM} for an estimate).
Consequently, until now results concerning the multi-scale statistical properties of
anisotropic fluctuations
have been characterized by considerable scatter,
thus calling into question their universal nature and in some instances even
jeopardizing the fundamental postulate of small-scale isotropy \cite{WS02}.

\noindent The main features of this work
are the following:
%%% changed BY KI due to Ref.1 comments
%First, we have achieved unprecedented Reynolds numbers for a paradigmatic 
%homogeneous shear configuration
%obtained by studying a random Kolmogorov Flow (RKF).
First, we have achieved sufficiently high Reynolds numbers 
for a paradigmatic 
homogeneous shear configuration
obtained from a random Kolmogorov Flow (RKF).
%Second, we have adopted a highly accurate Lebedev quadrature \cite{lebedev99}
%for expanding the
%correlation functions in the irreducible representations of the
%$\sot$ symmetry group (see Supplemental Material at \cite{SM} for details).
%The new $\sot$ algorithm overcomes the numerical bottlenecks
%in studying high Reynolds number flows by drastically decreasing
%the computational cost and thus expanding the range of
%problems where the $\sot$ decomposition is viable.\\
%%%KI changes due to Ref2
Second, we have adopted a highly accurate Lebedev 
quadrature \cite{lebedev75,lebedev76}
for expanding the
correlation functions in the irreducible representations of the
$\sot$ symmetry group.
On a $N^3$ grid, the new $\sot$ algorithm reduces the 
computational complexity  
from $\sim O(N^6)$ to $\sim O(N^3 \log N)$,
thus expanding the range of
problems where the $\sot$ decomposition can be viable
(see Supplemental Material at \cite{SM} for details,
also see \cite{dmitry2012,FFTW05}).
%%%%%%%%%%DNS parameters%%%%%%%%%%
\begin{table}
\begin{center}
\label{dns.tab}
\begin{tabular}{lcccccc} \hline
$\rel$ & $N$ & $\kres $ & $\kk_1$ &  $\kk_2$ & $F T_E/u'$ & $T_{\textrm{tot}}/T_E$ \\ \hline
$290$ & $1024$ & $1.94$  & $\pm(2,0,0)$ & $\pm(1,0,0)$ & $0.48$ & $118$ \\
$450$ & $2048$ & $1.92$  & $\pm(2,0,0)$ & $\pm(1,0,0)$ & $0.45$ & $50$ \\ \hline
\end{tabular}
\protect\caption{DNS parameters: Taylor scale Reynolds number $\rel = \sqrt{20 Re/3}$,
resolution $N^3$, $\kres = N\eta/3$ where
$\eta = (\nu^3/\meandiss)^{1/4}$ is the Kolmogorov length scale and
$\meandiss$ the mean dissipation, 
$\kk_1$, $\kk_2$ are the wave-vectors forced, $F T_E/u'$ is the
non-dimensional shear rate, where $F$ is the anisotropic forcing amplitude and
$T_{\textrm{tot}}/T_E$ is the length of the stationary state simulation in multiples of
large-eddy turnover time $T_E$.
\vspace{-5mm}
}
\label{dns.tab}
\end{center}
\end{table}
%%%%%%%%%%%%%%%%%%

\begin{figure}[!]
\begin{minipage}[t]{0.5\textwidth}
\hspace{-3.0em}
\includegraphics [width=0.42\textwidth]{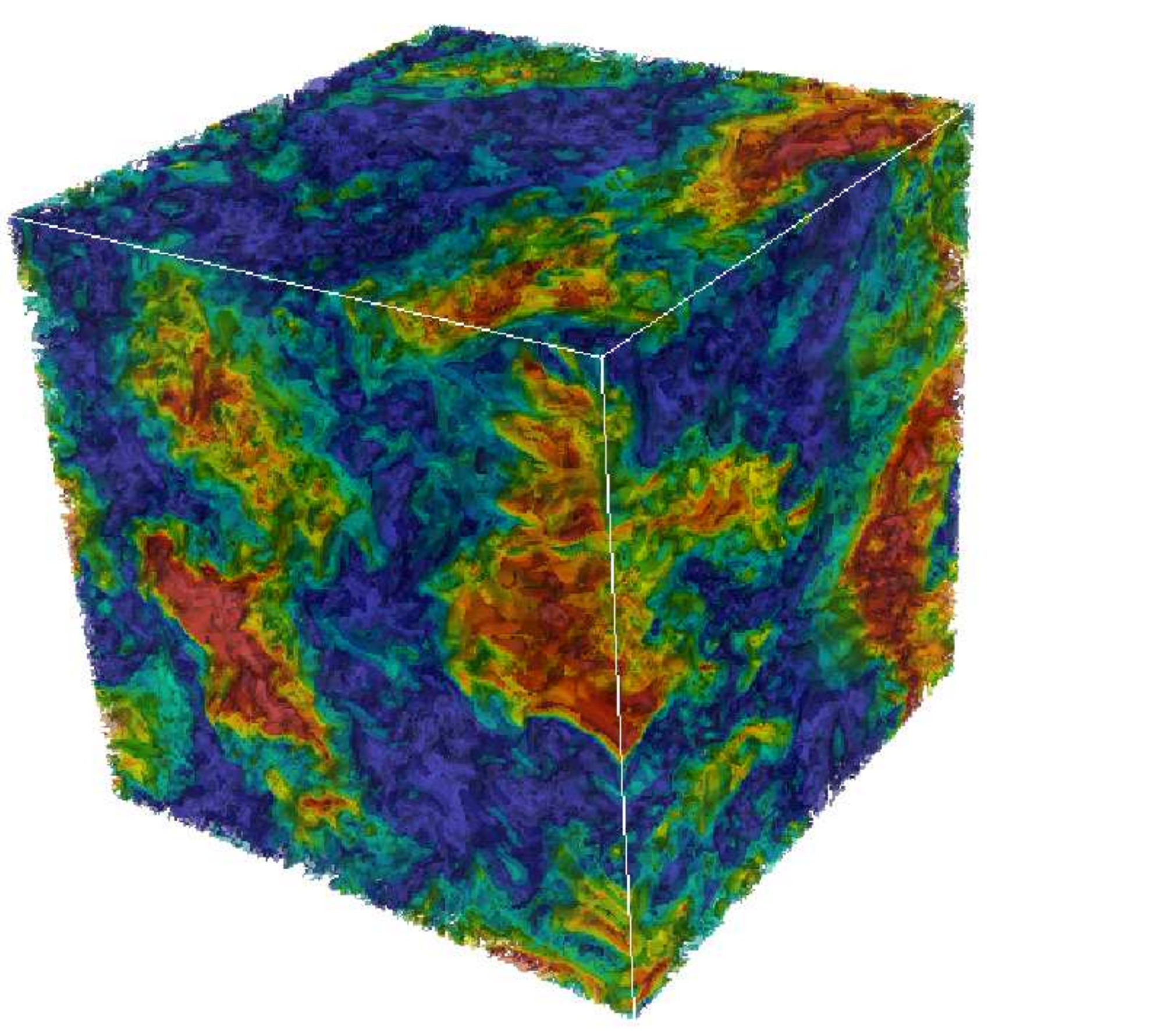}
\hspace{-2.5em}
\includegraphics [width=0.42\textwidth]{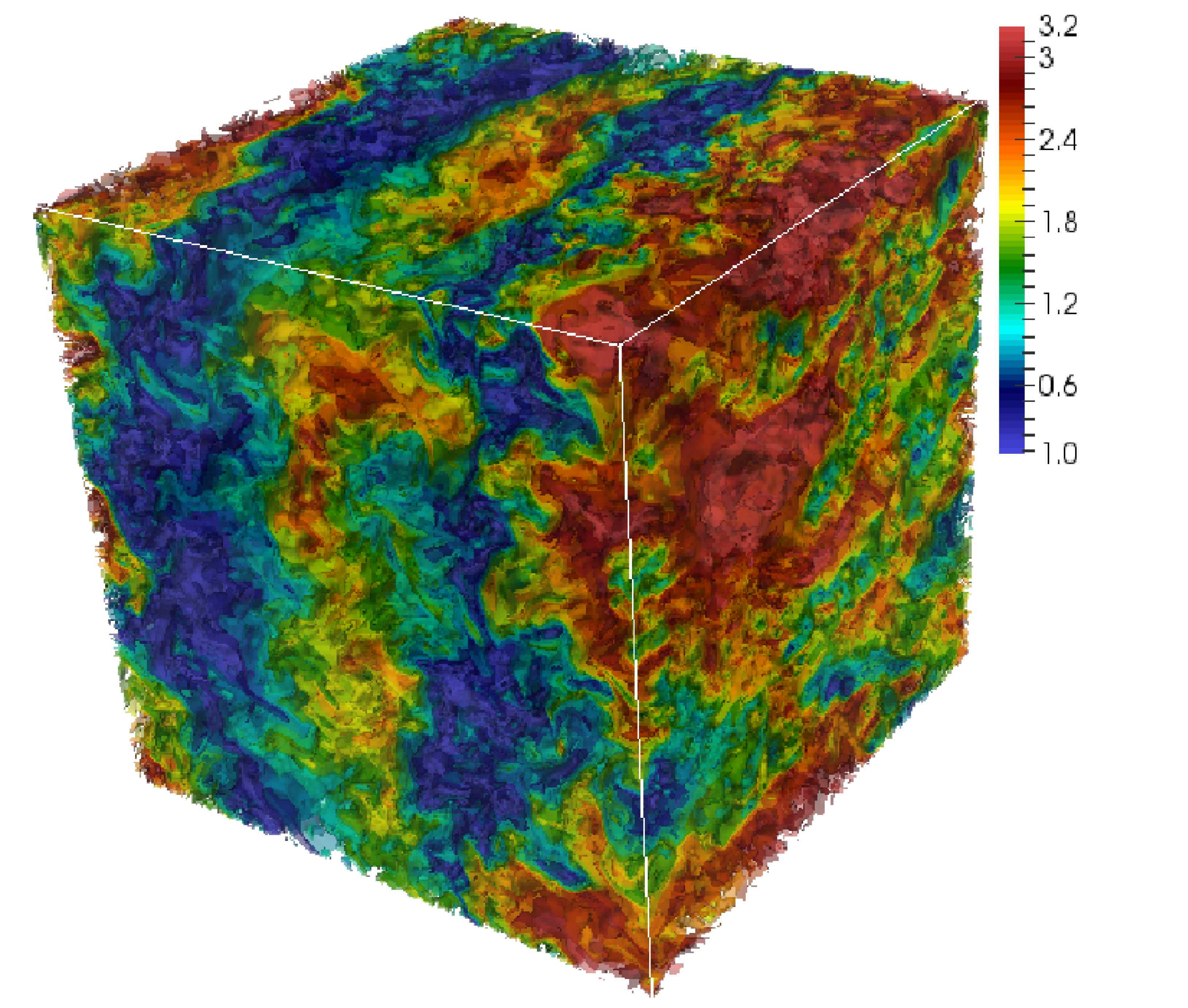} \\
\vspace{-7.8em}
\includegraphics [width=1.1\textwidth,right]{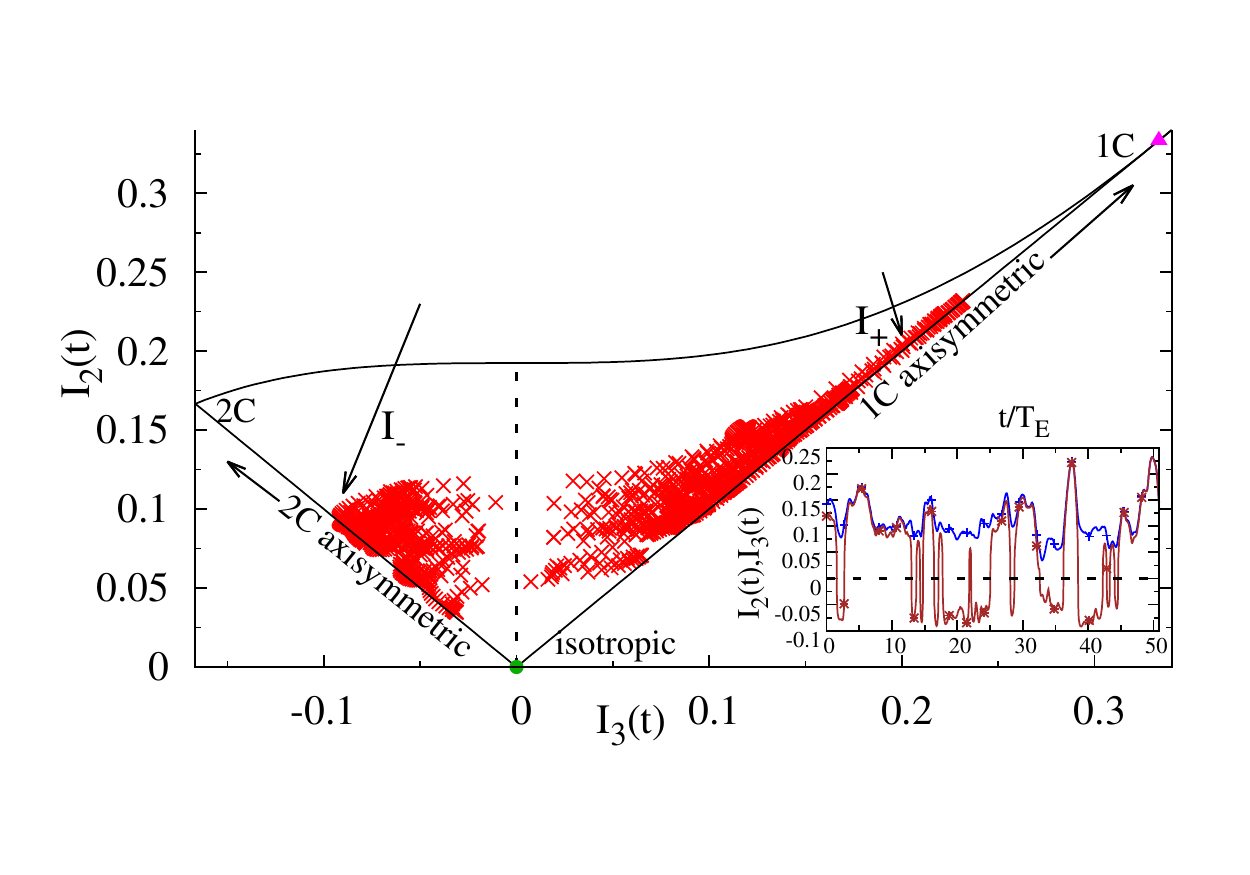}
\end{minipage}
\vspace{-14mm}
\protect\caption{Evolution of $I_3,I_2$ in the 
Lumley triangle ($\colr{\times}$) for the $2048^3$ RKF.
The laminar state given by the forcing configuration is shown by ($\colm{\blacktriangle}$). 
Inset shows steady-state evolution of $I_2$ $(\colb{+})$ and
$I_3$ $(\colbr{\ast})$ as functions of $\normt$.
Reference line at zero is the isotropic state.
Typical isocontours of the velocity magnitude in the
1C $(\iplus)$ and 2C $(\iminus)$ regions are also shown.
}
\label{i2i3lumley.fig}
\end{figure}

\noindent We discover that
the flow evolution reveals unexpected
bi-modal statistics of the energy containing scale,
characterized by chaotic oscillations between two states,
$\iplus$ and $\iminus$,
corresponding  to predominantly one-component (1C) and two-component (2C)
axisymmetric macrostates (see 3D rendering in Fig.~\ref{i2i3lumley.fig}), respectively. We
exploit the existence of the two macrostates
in assessing universality as a function of the large-scale flow configurations.
The main results are the following.
(i) By going to smaller and smaller scales, isotropy is recovered faster
than  previously thought.
We argue that this is due to the existence of non vanishing anisotropic contributions
from the $j=4$ sector (see below)
discarded or incorrectly measured in previous works \cite{KS00,WS02,BT2001}.
%(ii) We show for the first time that both longitudinal and transverse velocity increments
%posses similar scaling exponents even for the anisotropic components and we confirm the 
%theoretical expectation that 
%all non-universal contributions are hidden in the power-law prefactors, sector-by-sector in the
%$\sot$ decomposition.
%%%%%%%commented by KI in response to Ref. 1 
%%%%%%(ii) We show for the first time that  the anisotropic fluctuations
(ii) We show that the anisotropic fluctuations
of longitudinal and transverse velocity increments scale similarly.
We confirm the theoretical expectation that
all non-universal contributions are hidden in the power-law prefactors, sector-by-sector in the
$\sot$ decomposition of the velocity correlations.
%%%changed by KI due to Ref.1 comments
%%%(iii) We measure anisotropic scaling properties with unprecedented accuracy
%and find that contrary to previous expectations,
%the exponents do not saturate at higher orders, that they are universal and
%Reynolds independent at least up to the values investigated here.
(iii) The anisotropic scaling properties,
contrary to previous expectations based on low Reynolds number
calculations,
do not saturate at higher orders, that they are universal and
Reynolds independent at least up to the values investigated here.
%%%%%%%%%%%%

\noindent We study the
RKF \cite{BT2001} by evolving the 3D incompressible Navier-Stokes equations
in a tri-periodic domain,
\beq
\label{nse.eq}
\dho \uu/\dho t + \uu \cdot \nabla \uu = -\nabla p/\rho + \nu \nabla^2 \uu + \uf \;,
\eeq
where $p$ is the pressure and $\rho$ is the constant density.
An anisotropic and statistically stationary state is attained by forcing only
two wavenumbers (see Tab.~\ref{dns.tab}) in the $y$-direction,
%$f_y(\ux,t) \!  = \! F \sum_{\kk= \kk_1,\kk_2} \hat f_y(\kk,t) e^{i \kk \cdot \ux}$, 
$f_y(\ux,t) \! = \! F [\td{f_y}(\kk_1,t) e^{i \kk_1 \cdot \ux} \! + \!
\td{f_y}(\kk_2,t) e^{i \kk_2 \cdot \ux}]$, where $F$ is a constant amplitude and the Fourier coefficients 
$\td {f}_y(\kk_{1,2},t)$ follow independent divergence-less Ornstein-Uhlenbeck 
processes \cite{lbprx16}.
At variance with the standard non-homogeneous Kolmogorov Flow \cite{BO96},
we recover translation invariant statistics by averaging the 
random forcing phases over
multiple large-eddy turnover times, $T_E \equiv r_f/u'$.
%%%%%%%%%smust%%%% 
%%%%%%%%%%%%%%%%%%%%%%%
To quantify the anisotropy in the energy containing scales
we examine the temporal evolution of the two 
invariants $I_2(t) = (\bij b_{ji}/6)^{1/2},
I_3(t) = (\bij b_{jk} b_{ki}/6)^{1/3}$ of the  Reynolds stress, 
$\bij = (\la u_i u_j \ra/\la u_k u_k \ra) - \delta_{ij}/3$ \cite{choilumley01}. 
Figure \ref{i2i3lumley.fig}
shows $I_2$ plotted against $I_3$ at different
time instants, in the Lumley triangle \cite{lum78}. 
Surprisingly enough, despite the high 
Reynolds number, the dynamics is attracted by two different anisotropic
axisymmetric states $\iplus$ and $\iminus$  where $I_3 > 0$
and $I_3 < 0$ respectively.
The isocontours of the kinetic energy magnitude reveal a stark contrast in the
large scale structures between the 1C and the 2C macrostates (see Fig.~\ref{i2i3lumley.fig}).
The transition between $\iplus$ and $\iminus$ occurs suddenly during the time evolution, 
as shown in the inset of the same figure
by the temporal evolution of the two invariants. Notice that the large scale
configurations always avoid the
isotropic $I_2=I_3=0$ state.
The oscillations in the Reynolds stress
suggest the existence of multiple turbulent states akin to those found in
Taylor-Couette and Von Karman swirling flows \cite{Lohse14,SM14,SM15}.
Here, to assess the
degree of small-scale universality at changing the large-scale anisotropy 
we will show results
conditioned on the sign of $I_3$.\\
%%%%%%%%%%%%%%%%%%%%%%%%%%%%%%%%%%%%%%%%%%%%%%%%%%%%%%%
%%%%%%%%
%\noindent {\sc $\sot$ Decomposition}.~The longitudinal and transverse velocity 
\noindent The longitudinal and transverse velocity 
increments are defined as
$\dru_L (\ux,\ur) \equiv  \druvec(\ux,\ur)\cdot {\urh}$ and 
$ \druvec_T (\ux,\ur) \equiv \druvec(\ux,\ur)-\dru_L (\ux,\ur)\urh $ respectively, where 
$\druvec (\ux,\ur)  \equiv \uu(\ux+\ur)-\uu(\ux)$ is the two-point velocity difference
at separation vector $\ur$ and
$\urh$ is the unit vector along $\ur$.
The $p^{\textrm{th}}$ order
%longitudinal and transverse structure functions, abbreviated as 
%LSF and TSF respectively, are:
longitudinal structure function (LSF) and
transverse structure function (TSF), are
\beqa
\label{lsf.eq}
S^{(p,L)}(\ur) &\equiv& \la(\dru_L (\ux,\ur) )^p \ra \;, \\
\label{tsf.eq}
S^{(p,T)}(\ur) &\equiv& \la (\druvec_T (\ux,\ur) \cdot \druvec_T (\ux,\ur))^{p/2} \ra \;,
\eeqa
where $\la \cdot \ra$ denotes space and time averages. Since 
$S^{(p,L)}(\ur)$ are scalar functions of a vector arguments, they
can be expanded in spherical harmonics $\yjmr$ \cite{alp99} as,
\beq
\label{sflsh.eq}
S^{(p,L)}(\ur)  = \sum^{\infty}_{j=0} \sum_{m=-j}^{m=+j} S^{(p,L)}_{j,m}(\urmag) \yjmr \;.
\eeq
The index $j$ labels the different degrees of anisotropy, while the dependency on $m$ 
distinguishes different degrees of freedom within a given anisotropic sector.
The TSF can be expanded similarly via the corresponding projections
$S^{(p,T)}_{j,m}$.
The projection on the $j=0$ sector corresponds
to the isotropic case, the only one that will survive if the external forcing
is invariant under rotation.
Theoretical speculations 
suggest that at high enough $Re$ and for small enough 
scales $r \ll r_f$, a foliation of the physics in  different 
$j$-sectors occurs, characterized by different power law scaling
%\cite{kurien00,bp05},
\cite{kurien001,bp05},
%%%%%%%%%%%%%%%%%%%%%
\beq
\label{sf.eq}
\! \! \! \! S^{(p,L)}_{j,m}(\urmag) \! \!  =  \!\! 
\ljm \! \Big ({\frac{\urmag}{r_f}} \Big )^{\vslp} \! \! ;
S^{(p,T)}_{j,m}(\urmag) \!\!  = \!\! 
\tjm \! \Big ({\frac{\urmag}{r_f}} \Big )^{\vstp} \! \! \! . 
\eeq
%%%%%%%%%%%%%%%%%%%%%
%%%%%%%%%%%%%%%%%%%%%%%%%
\begin{figure}[!]
\begin{minipage}[t]{0.5\textwidth}
\vspace{-10mm}
\includegraphics [width=1.1\textwidth,center]{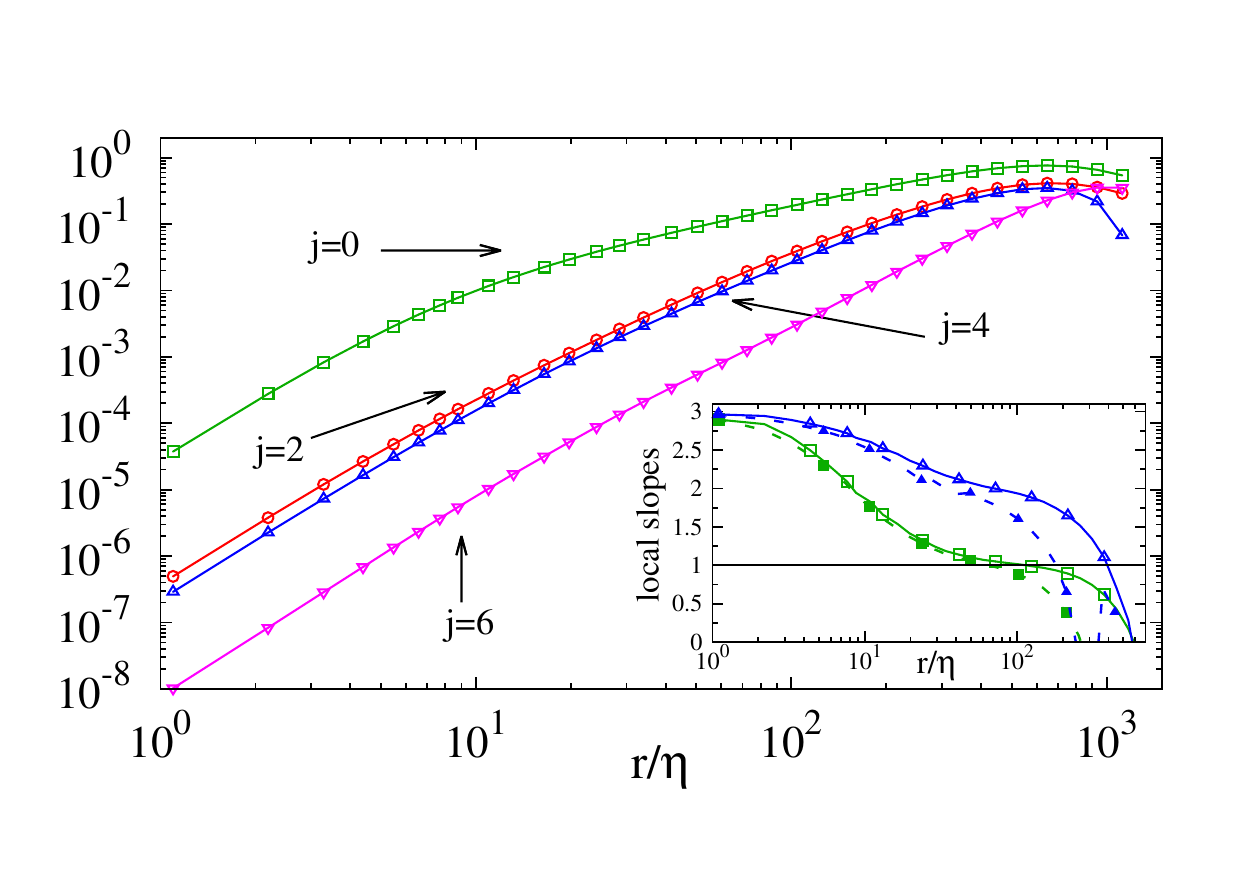}
\end{minipage}
\vspace{-11mm}
\protect\caption{Log-log plot of  $|S^{(3,L)}_{j,m}| $ vs $r$
for different  $(j,m)$ sectors
at $\rel=450$.
Symbols correspond to sectors 
$(0,0)$ $(\colg{\square})$,
$(2,0)$ $(\colr{\circ})$,
$(4,4)$ $(\colb{\triangle})$ and 
$(6,6)$ $(\colm{\triangledown})$, respectively. 
Inset shows corresponding logarithmic local slopes for sectors $(0,0)$ and
$(4,4)$ at $\rel=300$ (closed symbols) and $\rel=450$ (open symbols). Horizontal line
at $1.0$ is the exact result for sector $(0,0)$ \cite{K41b}.
}
\label{scaling.fig}
\end{figure}
%%%%%%%%%%%%%%%%%%%%%%%%%%%%%%%%%%%%%%%%%%%%%%%
All questions can then be translated in terms of the above defined quantities. 
Recovery of isotropy (universality) implies that a strict
hierarchy exists among the isotropic and anisotropic exponents,
$ \xi^{L}_0(p) < \xi^{L}_{j>0}(p) $.
The rate of recovery being measured by the gap 
between the exponents of the same order:
smaller the gap, slower the anisotropic contributions decay.
Theoretical considerations suggest that the 
exponents $\vslp$ and $\vstp$ are {\it universal}, i.e.~independent of the 
large-scale configuration.
The prefactors $\ljm$ and $\tjm$ must be {\it non-universal}
being determined by the matching for $\urmag \sim r_f$.
The exact expansion (\ref{sflsh.eq}) together with scaling 
assumption (\ref{sf.eq})
imply that in presence of anisotropy, multiple power laws are present 
in the undecomposed correlations such as $S^{(p,L)}(\ur)$ 
and hence non-trivial, sub-leading terms can contaminate their 
scaling behaviour. 
Conversely, the projected components $S^{(p,L)}_{j,m}$,
must show a pure power law behaviour.
In Fig.~\ref{scaling.fig} we asses the rate of recovery-of-isotropy by plotting
the magnitudes $|S^{(p,L)}_{j,m}(\urmag)|$
for $p=3$, up to $j=6$ (we omit those $(j,m)$ sectors that have negligible intensity or that 
have similar scaling properties).
All projections exhibit a clear power-law
behavior.
%%%KI comment for Ref2
The isotropic projection scales quasilinearly in 
the scale range $44 \le \normr \le 350$,
%as required by the $4/5$th law \cite{K41b} for asymptotic $\rel$ 
as it does in an isotropic flow, due to the $4/5$th law 
%for asymptotic $\rel$ 
%(see Refs.~\cite{qian97,lund02} for a discussion on finite 
%$\rel$ corrections)
\cite{K41b,qian97,lund02}.
All sectors have comparable magnitude at the forcing scale, 
confirming the strong anisotropy  of the energy containing scales.
In contrast, the anisotropic projections
become more and more sub-leading 
with decreasing $r$.
The quality of the scaling properties are shown in the inset of Fig.~\ref{scaling.fig}
where we  compare the logarithmic derivatives 
of $j=0$ and $j=4$ at two different Reynolds numbers.
Similar plots are obtained for other moments and for transverse increments (see also later). 
It is important to stress that the anisotropic projections shown in  Fig.~\ref{scaling.fig} 
display a quality of scaling never achieved before concerning both statistical accuracy and 
extension of the inertial range of scales.
An extremely high numerical and statistical accuracy is required to 
disentangle fluctuations that differ up to four orders of magnitude
(compare sectors $j=6$ and $j=0$ at the smallest $r$). 
These results have been possible due to the highly accurate quadrature 
that has been used for the $\sot$ decomposition (see Supplemental Material 
at \cite{SM} for details).
%%%%%%%%%%%%%%%%%%%%%%%%%%%%%%%%%%%%%%%%%%%%%%%%%%%%%%%%%%%%%%%%%%%%
\begin{figure}
\begin{minipage}[t]{0.5\textwidth}
\vspace{-10mm}
\includegraphics[width=1.1\textwidth,right]{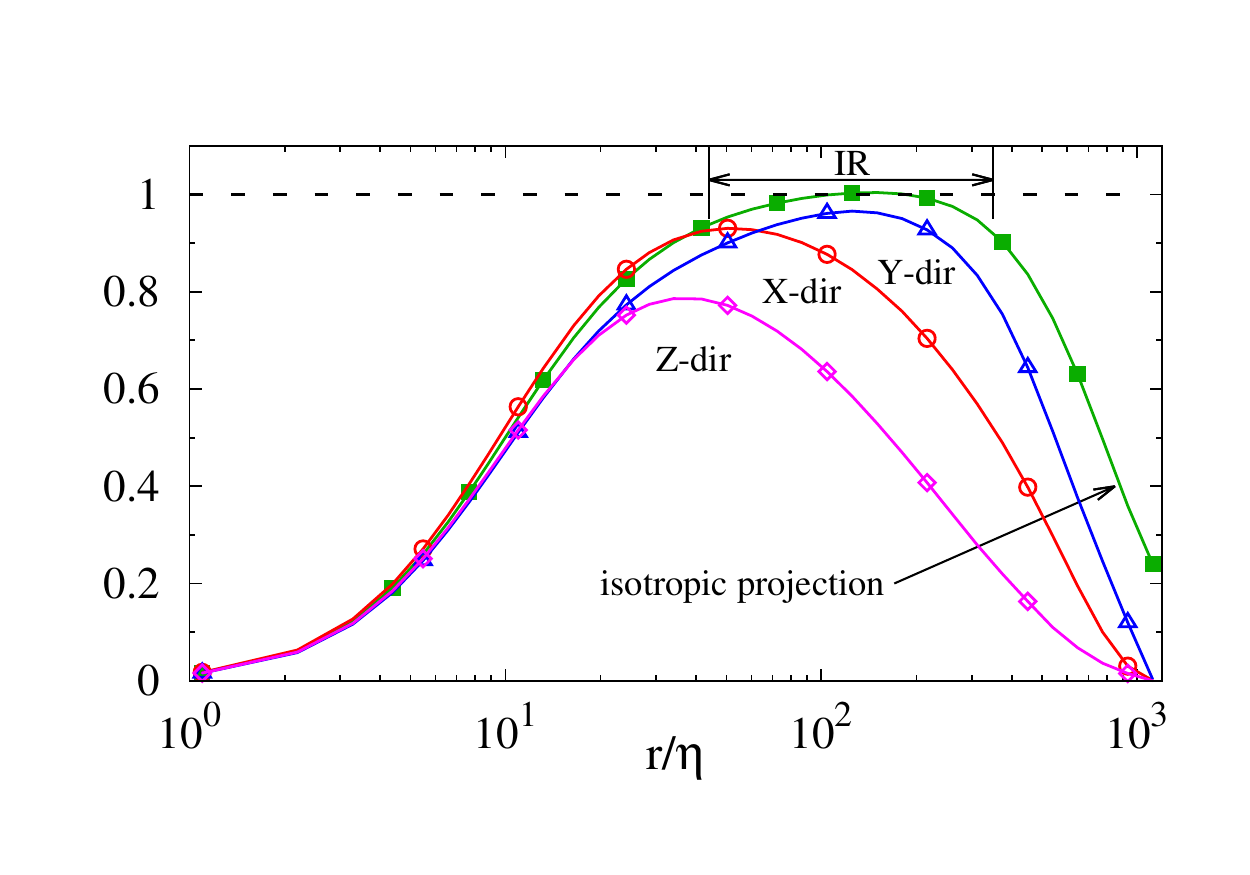}
\end{minipage}
\vspace{-11mm}
\protect\caption{Lin-log plot of undecomposed compensated 
structure function
$-5 S^{(3,L)}(\ur)/4 r\meandiss$
at $\rel = 450$ along the Cartesian directions, $\hat x$ 
$(\colr{\circ})$, $\hat y$ $(\colb{\triangle})$ and  $\hat z$ $(\colm{\diamond})$.
The corresponding projection on the isotropic sector, 
$-5 S^{(3,L)}_{(0,0)}(r) Y_{00}/4 r\meandiss$ is given by
$(\colg{\blacksquare})$. 
Inertial range (IR) is taken as that range within $5\%$ of the exact
IR result, shown by the dashed line at unity.
}
\label{dlll.fig}
\end{figure}
%%%%%%%%%%%%%%%%%%%%

\noindent Despite anisotropies being sub-leading
at the small scales, their cumulative effects are important and strongly influence
scaling laws if not properly decomposed.
This is shown in
Fig.~\ref{dlll.fig} which compares the 
undecomposed third order LSF along the three Cartesian directions along with the
projection on the isotropic sector, all compensated with 
the exact isotropic $4/5$th linear behaviour, $ -4/5 \epsilon r$. 
The undecomposed correlations do not compensate 
well and depend on the chosen direction.
In contrast, the isotropic sector confirms the K41 plateau \cite{K41b}
on a wide range of scales\citep{TKE03}.
%%%%%%%%%%%%%%%%%%%%%%
%%%%%%%%%%%%%%%%%%%%%%%%%
\begin{figure}
\begin{minipage}[t]{0.5\textwidth}
\vspace{-12mm}
\includegraphics [width=1.1\textwidth,right]{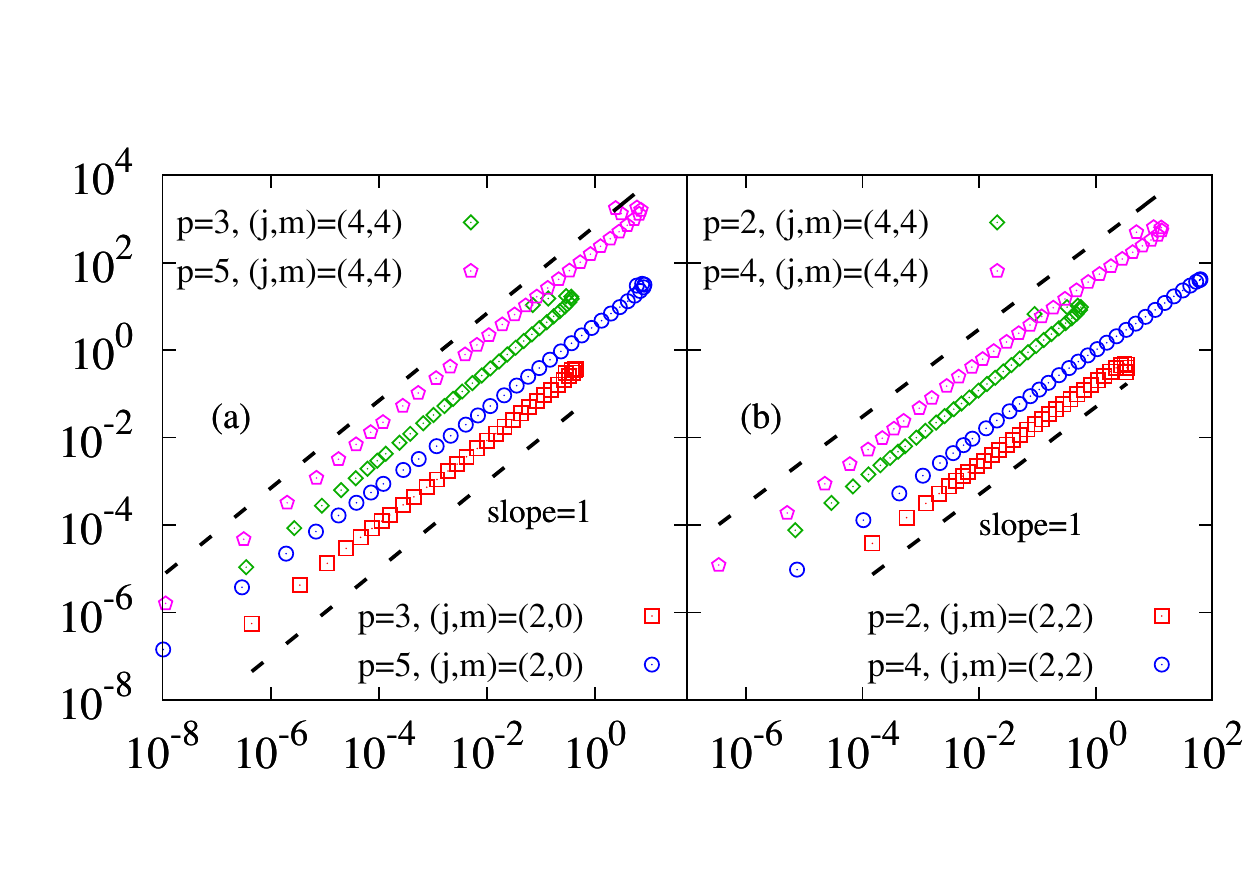}
\end{minipage}
\vspace{-10mm}
\protect\caption{(a): log-log plot of $S^{(p,L)}_{j,m}\big \vert_{\iplus}$  vs. $S^{(p,L)}_{j,m}\big 
\vert_{\iminus}$ at various orders $p$ and sectors $(j,m)$. (b): 
the same but for transverse structure functions. 
Universality with respect to large scale conditioning corresponds to slope $1$ as
shown by the dashed straight lines.
For clarity, curves are offset upwards.
}
\label{essaniso.fig}
\end{figure}
%%%%%%%%%%%%%%%%%%%%%%%%%%%%%%%%%%%%%%%%%%%%%%%%%%%%%%%%%%%
%%%%%%%%%%%%%%%%%%

\noindent To assess universality of the scaling properties sector-by-sector,
we show in Fig.~\ref{essaniso.fig} 
that both $S^{(p,L)}_{j,m}$ (left panel)
and $S^{(p,T)}_{j,m}$ (right panel) scale similarly when conditioned on 
$\iplus$ or $\iminus$ events.
Using a least-square fit we find that the relative
scaling exponents for all curves  is $\sim 1$ within $5 \%$.
%It appears clear that the relative scaling  are the same in the inertial range 
%for each given $j$-sector and for each moment $p$,
%while the overall intensity is strongly dependent on the $I_3$ value. 
This supports the foliation argument that the scaling exponents,
sector-wise are immune to anisotropic large scale effects and are hence universal. 
%%%%%%%%%%%%%%%%%%%%%%%%%%%%%%%%%%%
\begin{figure}
\begin{minipage}[t]{0.5\textwidth}
\vspace{-8mm}
\includegraphics [width=1.1\textwidth,right]{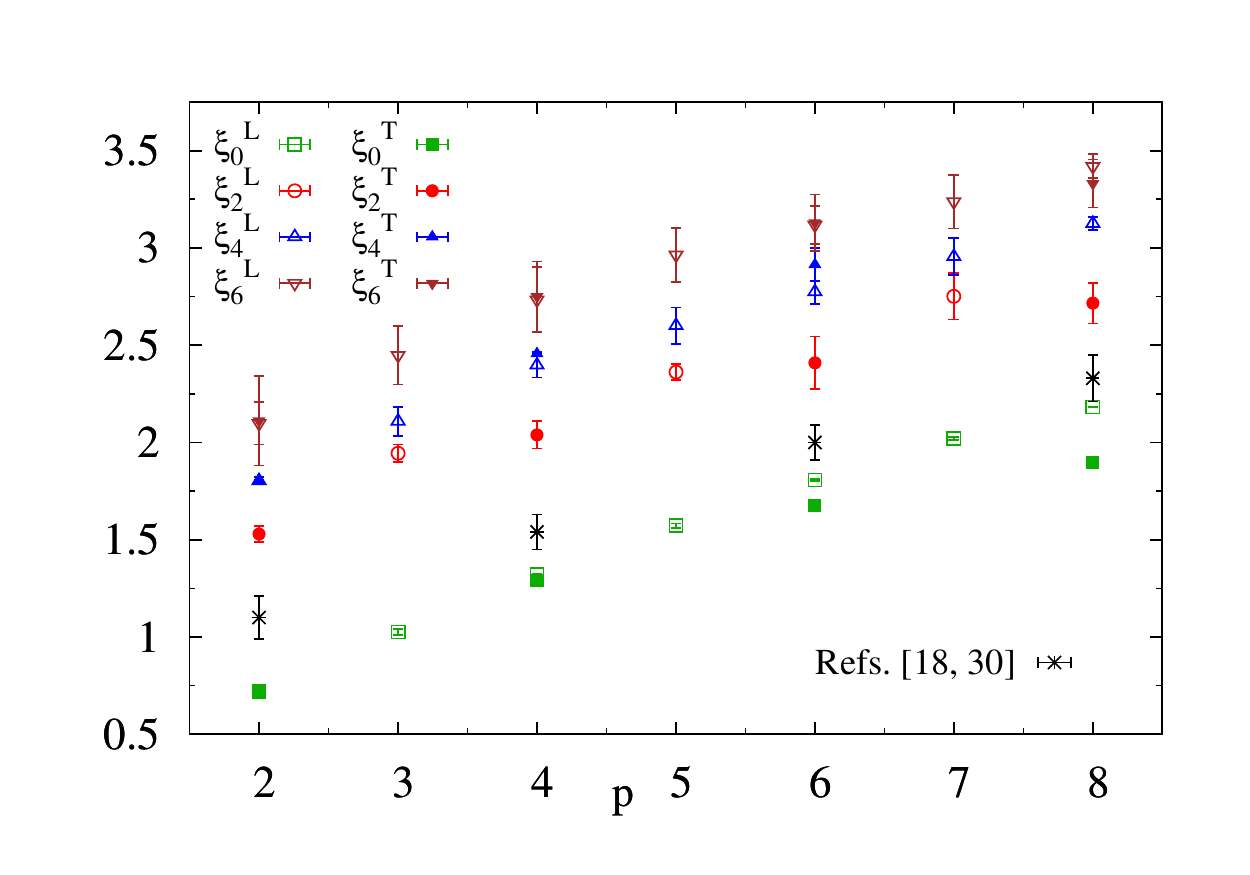}
\end{minipage}
\vspace{-9mm}
\protect\caption{Summary of
scaling exponents of $S^{(p,L)}_{j,m}$ (open symbols)
and $S^{(p,T)}_{j,m}$ (closed symbols)
{\it{vs}} order $p$
for sectors $j \le 6$.~Error bars indicate $95 \%$ confidence intervals.
Data from experiments for $j=2$ $(\ast)$
given for comparison \cite{KS00,WS02}.
}
\label{exp.fig}
\end{figure}
%%%%%%%%%%%%%%%%%%%%%%%%%%%%%%%%%%%%%%%%%%%%%%%%%%%%%%%%%%%%%%%%%%%%%%%%%%%%%%%%%%%%%%%%%%%%%%%%%%

\noindent The summary for all scaling exponents of different LSF and TSF 
projections
on the different sectors are plotted for various
orders in Fig.~\ref{exp.fig}. 
A few final comments are in order. 
(i) Both $\vslp$ and $\vstp$ have similar values,
except for small deviations at orders $p \ge 6$ in the $j=0$ sector
(see Refs.~\cite{krsht,benzi10,KI16} for a discussion on the Reynolds number 
dependency of the isotropic exponents).
(ii) At any given order $p$,
a finite gap exists between isotropic and anisotropic exponents, 
indicating a strict asymptotic recovery of isotropy
{\it{for the whole probability distribution function}}.
(iii) For any  $j$, the scaling exponents do not scale 
linearly in $p$, contrary to the dimensional prediction
$\vsdim = (p+j)/3$ proposed in 
Ref.~\cite{BDLT2002},
indicating that anomalous scaling is also present in $j>0$ sectors.
Importantly enough, 
the new $\sot$ scheme using a high order Lebedev rule 
\citep{lebedev99,jburk}
enables us to clean the previously reported results.
For example,
in contrast to Ref.~\cite{BT2001}, we find that
the exponents in any given anisotropic sector
increase with order $p$ with no apparent saturation.
%We have verified that the scaling exponents do not saturate at least 
%for sectors $j \le 6$ for $\rel \sim 100 - 450$. 
We contend that the saturation observed
in Ref.~\cite{BT2001} is
due to spurious effects induced by combination of 
poor accuracy in the $\sot$ expansion and potential contamination 
by hyper-viscous effects.
In Fig.~\ref{exp.fig} we also report 
results for the $j=2$ sector from the few prior experiments \cite{KS00,WS02}.
In experiments, it is difficult to perform measurements along
a sufficiently large number of directions
to adequately resolve the anisotropic 
fluctuations on the $2$-sphere, in contrast $\sim O(1000)$
different directions were used in this work.
As a result, experiments must resort to a fit for the entire
right-hand-side of 
Eq.~\ref{sflsh.eq} using data along a few directions only.
In order to reduce the number of fitting parameters the sum on all sectors 
is typically cut at $j=2$,
something that is clearly not enough in view of the 
results shown in Fig.~\ref{scaling.fig}. Indeed, 
we find that sector $j=4$ is almost as 
energetic as $j=2$, with a very similar scaling exponent, 
i.e.~the $j=4$ contribution is as important as $j=2$, at almost all scales.
%As one can see from Fig.~\ref{exp.fig}, our exact decomposition shows a 
%clear difference with the 
%fit for $j=2$ of Refs.~\cite{KS00,WS02}.
%As one can see from Fig.~\ref{exp.fig}, results from
%%%changed by KI for Ref. 1
Figure \ref{exp.fig} shows that the results from 
our exact decomposition
clearly differ with that of Refs.~\cite{KS00,WS02}, wherein 
sectors $j \ge 4$ are neglected.
%In the presence of many anisotropic sectors, any fit based only on one 
%anisotropic power law might be strongly affected by spurious cancellations, 
%explaining the large difference between our exact projection and the 
%empirical 
%experimental fit.\\
%%%%%%%%%%%%%
%%%changed by KI for Ref. 1
In the presence of many anisotropic sectors,
obtaining scaling exponents by assuming that only the lowest anisotropic
sector is dominant can strongly affect the measured rate of 
return owing to spurious cancellations.
Only the exact $\sot$ expansion allows the measurement of $\xi_{j}(p)$, 
devoid of contamination from sectors $j' \neq j$, thus
yielding a true gauge of the rate of return at a
given order $p$.  It remains to be clarified if in the homogeneous shear 
case analyzed in Refs.~\cite{SW2000,WS02} the scaling properties of 
high order sectors $j\ge 4$ are also as important as in the RKF.\\
%%%%%%%%%%%%%
%changed by KI due to Ref1
%%%%\noindent {\sc Conclusions}. We have used a combination of 
%%%%state-of-the-art DNS and an improved algorithm for the
%%%$\sot$ decomposition, to study anisotropy in high Reynolds numbers Kolmogorov Flows.
%%%%%%%%%
\noindent In conclusion, we have used an efficient
algorithm for the
$\sot$ decomposition, to study anisotropy in high Reynolds numbers Kolmogorov Flows.
We have found that the RKF develops a 
two-state attractor characterized by very different anisotropic large scale contents.
We have shown that the
scaling exponents in RKF are immune to
different large scale effects and hence are universal. The
magnitude of the anisotropic exponents
indicate that  isotropy is recovered at a faster rate than previously thought. 
Nevertheless,
projection on the $\sot$ is mandatory to detect a clean scaling,
since power laws exists only sector-by-sector.
We do not observe saturation of exponents at the higher $j$-sectors, 
indicating that intense anisotropic fluctuations are dominated by 
more than one singular structure.
Differently from  previous observations, we demonstrate that it is 
mandatory to resolve at least up to sector $j=4$ to have clean
scaling properties, 
sector by sector. We hope our study will stimulate further theoretical or 
phenomenological efforts to 
predict the scaling properties for all  $j$-sectors. 
%%%%%%%%%%%%%
It will be  important to extend this analysis to other turbulent flows, 
such as those in the presence of rotation, mean shear and magnetic field,
in order to establish, on a firmer basis the degree of universality.
The improvement provided by the {\it fast $\sot$ solver} opens
the road to perform such studies.
%%%%%%%%%%%%%%%%%%%%%%%%%%%%%%%%%%%%
%%%%%%%%%%%%%%%%%%%%%%%%%%%%%%%%%
%%%%%%%% END %%%%%%%%%%%%%%
\section{Acknowledgments}
We thank Susan Kurien for useful discussions.
We acknowledge  funding from the European Research Council under the European
Community’s Seventh Framework Program, ERC Grant Agreement No $339032$. 
The computations were performed using
resources provided through the PRACE initiative $\textrm{Pra}11\textrm{\_}2870$
at the CINECA Consortium.
We acknowledge the European COST Action MP1305 ``Flowing Matter". 
%%%%%%%%%%%%%%%% THE END %%%%%%%%%%%
%%%%%%%%%%%%%%%%%%%%%%%%%%%%%%
\bibliography{zebib}
%%%%%%%%%%%%%%%%%%%%%%%%%%%%%
\end{document}